\documentclass[runningheads]{llncs}
\usepackage{cite}
\usepackage[cmex10]{amsmath}
\usepackage{algorithmic}
\usepackage{array}
\usepackage[pdftex]{graphicx}
\usepackage{textcomp}
\usepackage{latexsym}
\usepackage{mathptmx}
\usepackage{float}
\RequirePackage{accsupp}
\usepackage{pdfcomment}
\usepackage{comment}
\usepackage[autostyle]{csquotes}

\author{Philip Feldman\inst{1, 2} \and
	Aaron Dant\inst{1} \and
	Wayne Lutters\inst{2}}
\authorrunning{P. Feldman et al.}
\institute{ASRC Federal, Columbia Maryland, USA \and
	University of Maryland, Baltimore County, Maryland, USA}

\begin{document}
	\title{Disrupting the Coming Robot Stampedes: Designing Resilient Information Ecologies}
	\maketitle
	
	\begin{abstract}
		Machines are designed to communicate widely and efficiently. Humans, less so. We evolved social structures that function best as small subgroups interacting within larger populations \cite{cucker2007emergent}\cite{miller1956magical}. Technology changes this dynamic, by allowing all individuals to be connected at the speed of light. A dense, tightly connected population can behave like a single agent. In animals, this happens in constrained areas where stampedes can easily form. Machines do not need these kinds of conditions. The very techniques used to design best-of-breed solutions may increase the risk of dangerous mass behaviors among homogeneous machines. In this paper we argue that ecologically-based design principles such as the presence of diversity are a broadly effective strategy to defend against  unintended consequences at scale. \cite{kirilenko2017flash}
	\end{abstract}

\section{The bonfire of the vanity plates}
The Southern California summer of 2024 is once again a scorcher. The record temperatures are mitigated by the cool air conditioning of the autonomous vehicles packing the freeways. They move fluidly, as traffic jams are generally a thing of the past. Sophisticated routing algorithms keep traffic flowing smoothly around any problems. On their designated lanes vehicles travel in tight, aerodynamic packs at inhuman speeds. Commute times that used to take hours now take minutes. The process is so efficient and trusted that passengers rarely even glance up from their screens.

A wildfire is rapidly descending on Los Angeles. Blown by gale-force winds, it covers 10 miles every 30 minutes, endangering the 405 near Bel-Air. Police shut down the highway, and the traffic quickly routes around the obstruction. Many of the alternate routes are also blocked, but the A* algorithm \cite{hart1968formal} is relentless in finding new routes. A solution is quickly found. Unfortunately, it is through burning neighborhoods not yet identified on the satellite map updates. Thousands of cars optimally converge on a two lane road, twisting along a ridge into the fire.

Although these cars are state-of-the-art and handle many circumstances flawlessly, they have not been trained to recognize sheets of fire blowing across the roads, generating temperatures that melt metal. To their sensors, the way ahead is clear.

Thousands of guidance systems agree that this is the best route. 

Thousands of cars charge into the flames.

As they enter the fire, antennas burn off, sensors malfunction and vehicles disappear from the network. Like stampeding buffalo, they swerve off the road and into the ravine. This happens \textit{fast}, with between 3-4 cars leaving the road \textit{per second} \cite{caltrans_2016}.

For the remaining cars fleeing the closed freeway, the way still appears open. Cars continue to flood through the gap until the pile of burning wrecks is large enough that the system finally recognizes the obstruction and searches for a new alternative, thankfully away from the inferno.

It takes weeks to figure out what happened and issue a patch to the millions of identical vehicles. They are back on the roads but just as vulnerable to the next unforeseen problem. Is there anything that can be done to change this calculus so that every potential does not have to be predicted?

\section{Understanding ecologies of artificial systems} 
Best design practices are about building for scale. These systems can be thought of as collections of mass-produced items operating as individual components or combined into a predictably functioning whole. But when the whole is an ad-hoc network of millions of independent components, these practices will no longer be sufficient to deal with emergent and unpredictable problems. For answers, we can look to ecosystems, where emergence and surprise are the bedrock of resilience \cite{holling2002resilience}.

Our work focuses on the behavior of individual agents and groups in \enquote{belief spaces}, the subset of information space that is associated with opinion (e.g., design, politics, \& pertinence). Our simulations explore mechanisms for how consensus develops, starting with high-dimensional concepts that are then simplified to produce alignment within groups \cite{DBLP:journals/corr/abs-1804-05409} \cite{DBLP:journals/corr/abs-1804-02251} \cite{Feldman:2018:MMS:3176349.3176353}. The type of interaction that emerges is a function of agent similarity combined with connection density and \enquote{stiffness}. Think of a noodle. Dry, it behaves like a single entity because every element is stiffly connected to every other. Cooked, the connections relax, and it is possible to move one end of the noodle without affecting the other. Applied to communication networks, it is easy to see that a densely connected network of identical agents communicating at lightspeed will behave very differently from a pre-industrial society communicating through letters and personal contact. The former has the capability and predisposition to behave as a monolithic entity, while the latter will have to adopt very different strategies to achieve their goals \cite{kiesler1992group}.   

The combination of lower-dimension belief space containing a closed, well connected network of similar individuals creates the conditions for stampede. The agents begin to behave as a block instead of a group. Social pressure is powerful, as group consensus can outweigh environmental factors \cite{grunbaum1998schooling}. In people, this manifests as runaway echo chambers. In machines, this can look like a stock market \enquote{flash crash}. 

These network properties become a set of \enquote{knobs} that can be used to affect the formation of group behavior. Our research indicates that a small number of \enquote{diverse} agents, injected into the awareness of a stampeding population can have beneficial effects. They enlarge the belief space so that the stampeding agents can be nudged away from their current trajectory. They slow the stampede, since more computation is required to evaluate the belief environment. They can peel away less rigid members, reducing the impacts when the stampede hits an environmental limit. 

Unfortunately, our simulations show that turning these knobs in the other direction can \textit{create stampedes}. It has been shown that social media can be used to increase polarization along \textit{emergent} poles \cite{giles2016handbook}. Examples have been documented of troll farms simultaneously organizing opposing groups \cite{stewart2018examining}. In these cases, bad actors are changing the underlying network structure to more densely connect like-minded people and push them further in a direction that they are already headed.

Diversity injection into social media feeds has the potential to disrupt artificial stampede conditions. In human systems this could take the form of adding random, interesting, known good information to the top of a news feed. For intelligent machines, using legal mechanisms similar to antitrust would ensure a minimum number of independent systems are incorporated in large scale applications like transportation. By enforcing diversity, we begin to gain the resilience to unpredictable threats as found in ecosystems.

\section{Evolved vs. designed systems} 
\begin{figure}
	\centering
	\begin{minipage}{.5\textwidth}
		\centering
		\fbox{\includegraphics[height=16em]{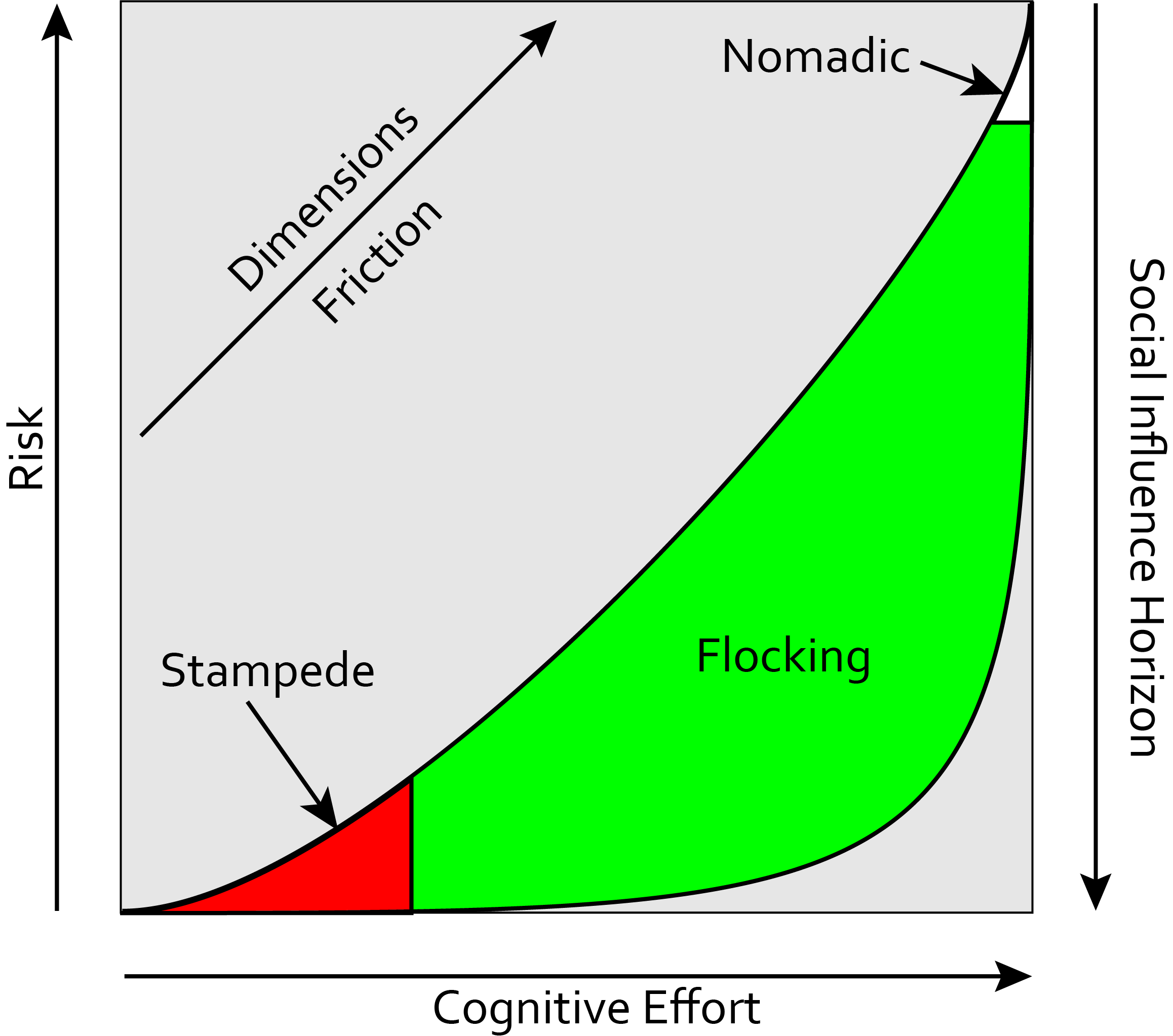}}
		\caption{\label{fig:N-F-S} Evolved systems}
	\end{minipage}%
	\begin{minipage}{.5\textwidth}
		\centering
		\fbox{\includegraphics[height=16em]{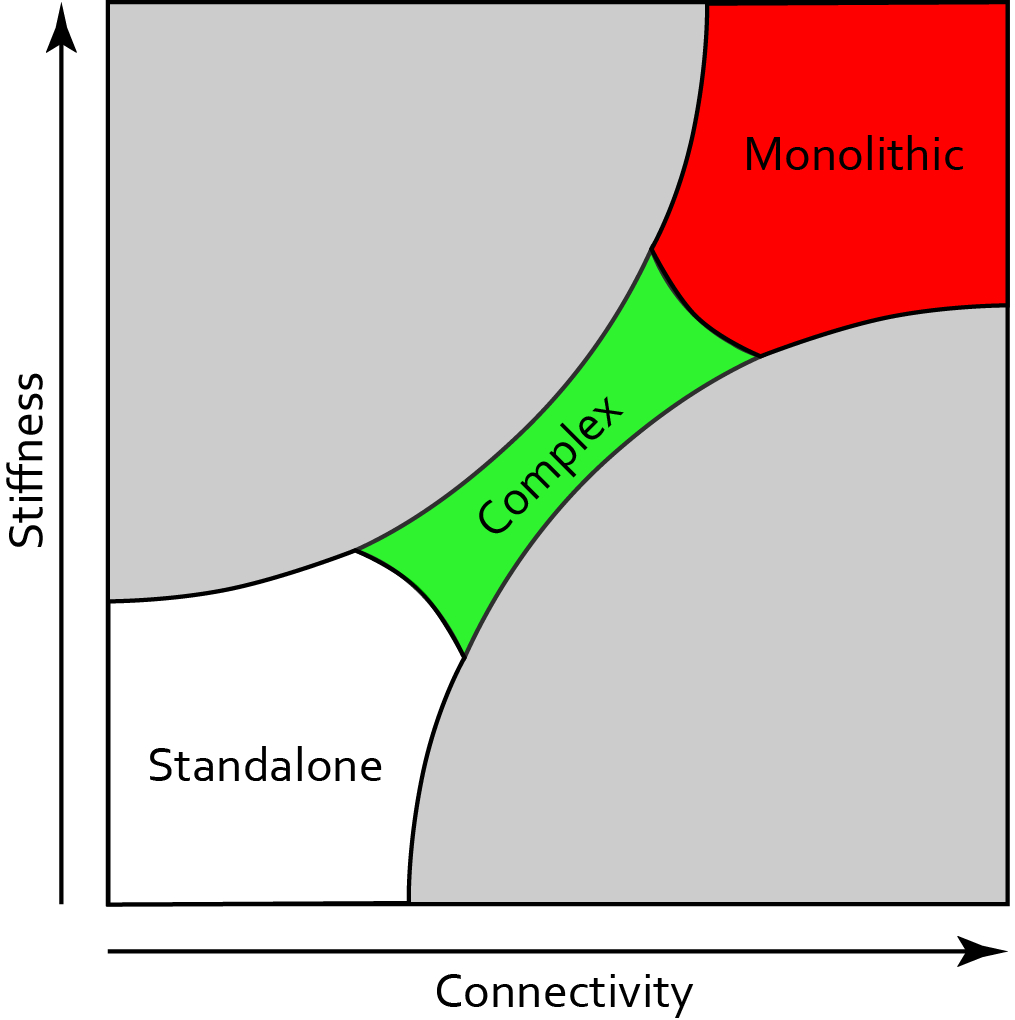}}
		\caption{\label{fig:Monolithic_complex_standalone} Designed systems}
	\end{minipage}%
\end{figure} 

Our work indicates that agents interact in one of three patterns (figure \ref{fig:N-F-S}). These phases emerge in evolved systems, and they have an equivalent in designed systems. These phases can be induced by changing the social influence horizon (SIH) of the agents, ranging from \enquote{stiff} to \enquote{loose}. These phases are:

\textit{Nomadic / Standalone} - Each agent is autonomous, and navigates belief space uninfluenced by other agents. Nomadic behaviors are risky for the individual but advantageous for the population, as they explore broader, possibly dangerous spaces. If they encounter something beneficial, they can bring it back to the population.

\textit{Flocking / Complex} - Agents' heading and velocity are affected by neighbors who are within their SIH. There is alignment between them, but the network is sparse and loose so that group orientations change over time. Flocking activity is a balance between environmental and social constraints.

\textit{Stampede / Monolithic} - All members are stiffly connected. Alignment can become total and supports runaway conditions \cite{lande1981models}. Such strongly aligned systems work well when there are few environmental risks but have difficulty in adapting to new situations \cite{grunbaum1998schooling}, and can literally run off cliffs \cite{patent2006buffalo}.

The middle ground between standalone and monolithic behavior is rarely touched by our designs. The systems that we develop operate at either end of this spectrum (figure \ref{fig:Monolithic_complex_standalone}). At the standalone end is NASA's Curiosity rover. At the other is a fully automated robot factory. We are starting to explore a middle ground of resilient, self-organizing systems, such as High Frequency Trading which coordinates using markets \cite{mackenzie2014sociology}, and Deep Neural Networks, where neuron \enquote{agents} interact using backpropagation to recognize patterns.  

\section{Engineering better information ecologies}
Setting up ecologies is very different from designing systems. Ecologies require a substrate that also serves as some kind of fitness test  \cite{kaufman1995home}. In living systems, this is typically energy (food \& light) and protection (defenses \& group safety). The fitness test is survival and reproduction. In designed systems, the substrate is us. The fitness test is sales and contracts, as experienced by companies, their managers, designers, and customers. Which is why \enquote{adaptation} consists of patches, upgrades, and new models. When we are part of the system, it is adaptable. When we are not, it withers and fades.

Designing resilient systems, using the principles of ecology, is a capacity that we need to develop  as we drift from our barely manageable current state towards more chaotic and unpredictable technological interactions. But there is some low hanging fruit. In particular, diversity injection. Let us revisit our firestorm example, but this time with an ecosystem of multiple different vehicles.

We are in the first group of vehicles on the same ridge road heading into the fire. In front of us, a handful of cars are already off the road, engulfed in flames. But the next car in line has frozen in the middle of the road. Its electronics are shorting out. The motor has seized, and the radio is blaring. This model is notorious for failures, and has been known to hit pedestrians. It is, in short, terrible. 

But it is \textit{different}. And it is \textit{blocking the road}.

As before, but much earlier, the routing algorithm kicks in and cars fleeing the freeway are redirected to places that are \textit{not} on fire. Resulting in hundreds of lives saved. All because there were a few, cheap, poorly made autonomous vehicles in the mix.

A population that is diverse has a greater chance of responding to unanticipated problems in unanticipated ways. The trade-off is lower efficiency, but that's a feature, not a bug. We need to start creating self-regulating systems that are inherently less likely to stampede out of control. Diversity injection is an accessible starting point. To begin this, \textit{we ourselves} need diversity in the form of transdisciplinary research. For example, useful frameworks may emerge from the interplay of information science, ecology, and economics. This will enable new design languages for safe, self-regulating systems. Without which, we run the risk of being victims of our own complex systems as they drive us unaware into the wildfire.

\bibliographystyle{splncs04}

\end{document}